\documentclass[aps,pre,showpacs,preprintnumbers,amsmath,amssymb,superscriptaddress]{revtex4}

\usepackage{graphicx}
\usepackage{bm}
\begin{document}
\title{Implementation of the histogram method for equilibrium
statistical models using moments of a distribution}
\author{Gabriel P\'erez$^*$}
\affiliation{Departamento de F\'{\i}sica Aplicada, Centro de Investigaci\'on y de Estudios Avanzados \\
del Instituto Polit\'ecnico Nacional, Unidad M\'erida \\
Apartado Postal 73 ``Cordemex'', 97310 M\'erida, Yucat\'an, M\'exico}
\date{\today}
\begin{abstract}
This paper shows a simple implementation of the Histogram Method for
extrapolations in Monte Carlo simulations, using the moments of the operators
that define the energy, instead of their histogram. This implementation is
suitable for extrapolation over several operators, a type of calculation
that is hindered by computer memory limitations.
Examples of this approach are given for the 2-D Ising model.
\end{abstract}
\pacs{PACS numbers: 02.70.Uu, 05.10.Ln, 05.50.+q}
\maketitle

\section{Introduction}

The calculation of the expectation value of some operator $\Phi$ in 
a system in contact with a heath-bath of temperature $T$
\begin{equation}
\langle \Phi \rangle_T = \frac{\sum_i{\Phi_i \exp(-E_i/k_B T)}}
{\sum_i{\exp(-E_i/k_B T)}},
\label{expectval}
\end{equation}
is the usual task of equilibrium statistical mechanics. 
Here $E_i$ and $\Phi_i$ are, respectively, the energy and the value of 
$\Phi$ associated to the $i$-th configuration, and the sum runs over 
all possible configurations.
These expectation values are usually
impossible to calculate analytically,
but there are many techniques to approximate expectation values in 
the canonical ensemble. Some of them are quite powerful and general, 
like high-temperature expansions and renormalization group techniques.

A different approach, quite successful in many 
applications, is the numerical simulation (``Monte Carlo Simulation'') 
of the ensemble \cite{metropolis1,newman-barkema,landau-binder}.
This paper is concerned with the Histogram Method, introduced 
years ago 
to extrapolate the results of a Monte Carlo simulation conducted
at a point in parameter space ---for instance, some given temperature 
and magnetic field---, to a range of those parameters. 
It was formulated by by Ferrenberg and Swendsen 
\cite{ferrenberg1,ferrenberg2}, 
although there are some earlier proposals \cite{valleau1}. 
For completeness, a short description of this technique follows:
Assume for instance that one has a model where the only
control parameter is the temperature, and wants to compute the behavior of 
some function $f$ of the 
energy $E$. At some given $T$ one performs a
Monte Carlo simulation using any given algorithm that generates $Q$ 
configurations 
with the correct probability, given by their Boltzmann weight, 
measures $f(E)$ for each configuration, and calculates finally 
the average value $\langle f(E) \rangle$. 
This average may be taken directly, or going through the preliminary 
construction of a normalized histogram $W_T(E)$ for the energies found 
in the simulation. One can write then
\begin{equation}
\langle f(E) \rangle_T  
\approx \frac{1}{Q} \sum_{q=1}^Q f(E_q)
= \sum_E {W_T(E) f(E)},
\label{gethistogram}
\end{equation}
which is approximation to Eq.~(\ref{expectval}) with $\Phi = f(E)$. 
This equation can also written in terms of the density of states $g(E)$
\begin{equation}
\langle f(E) \rangle_T = \frac{\sum_E g(E) f(E) \exp(-E/(k_B T))}
{\sum_E{g(E) \exp(-E/(k_B T))}}.
\label{density}
\end{equation}
Comparing Eqs.~(\ref{gethistogram}) and~(\ref{density}) 
it is clear that $W_T(E)$ is proportional to $g(E) \exp(-E/(k_B T))$, 
and so an approximation for the density of states ---ignoring 
normalization for the moment--- is given by
$g(E) \approx W_T(E) \exp(E/(k_B T))$. 
Notice that $g(E)$ is a property of the system itself, independent of 
temperature.

With an approximated density of states at hand the results of the
simulation are extended to other temperatures. Eq.~(\ref{density}) written
for a different temperature $T^\prime$ gives
\begin{eqnarray}
\langle f(E) \rangle_{T^\prime} &=& \frac{\sum_E g(E) f(E) \exp(-E/(k_B T^\prime))}
{\sum_E{g(E) \exp(-E/(k_B T^\prime))}} \nonumber \\
& \approx &
\frac{\sum_E W_T(E) f(E) \exp(E/(k_B T) - E/(k_B T^\prime))}
{\sum_E{W_T(E) \exp(E/(k_B T) - E/(k_B T^\prime))}},
\end{eqnarray}
and one says that the histogram $W_T(E)$ has been {\em reweighted\/}. 

\section{Expansion in Moments}

In general, with the Histogram Method results 
obtained at a given point in the space of control parameters can 
be extrapolated to a neighborhood of that point. However, the
implementation of the method can be
cumbersome for models with more than one parameter, because
the size of the histograms generated becomes too large. 
For instance, an Ising model in a simple cubic lattice, 
with nearest-neighbor, next-nearest-neighbor and magnetic couplings 
has a 3D parameter space.
When applying the histogram method to this system
one faces the problem of handling a histogram
with a total number of bins that rapidly becomes impossible to 
accommodate in the computer's physical memory. For instance,
for a very modest size $L = 16$, one needs to store some
$16^3 \approx 4 \times 10^3$ possible values for 
each of the 3 couplings, and this gives a histogram with around 
$64 \times 10^9$ bins. Working with 4 Byte integers (which may be 
insufficient for very large runs) gives a
memory requirement around 250 GBytes, realizable in large 
workstations and supercomputers, but usually beyond the reach of the
commodity machines used by many scientist today \cite{not-dense}. 

Among possible solutions one could coarse-grain the binning of 
the histogram (this is unavoidable if one is dealing with continuous 
variables, like in the {\em XY} or Heisenberg Models); 
one could also store a record of the operators of interest for all configurations 
generated by the Monte Carlo algorithm (this has been
recommended for Hamiltonians with continuous variables 
\cite{newman-barkema,ferrenberg3}). 
Here I propose a different option that arises from two simple facts: 
one, most of the 
quantities one is interested in calculating can be expressed in terms 
of moments of the same operators that constitute the Hamiltonian.
Two, the reweighting 
function can be easily expanded in a power series. Then, the physical 
quantities one wants to estimate at 
different values of the couplings can be obtained as power series in 
the increments of said couplings, using the moments of the operators 
obtained at some fixed point in parameter space as
coefficients. A similar approach was proposed in Ref.~\cite{rickman1}, 
using cumulants instead of moments; This approach via moments is bit simpler. 
Here I also show some details of implementation needed to insure 
larger ranges of numerical convergence.

To fix ideas, I shall use a simple example, namely, the Ising Model with 
zero magnetic field. The Hamiltonian is
\begin{equation}
{\cal H} =  -J \sum_{<a,b>} S_a S_b. 
\label{ising2d}
\end{equation}
Assume now that one calculates the expectation value of
some power of the magnetization, say $\langle M^2 \rangle$, at some given
temperature $T$. Define for convenience the adimensional energy operator 
$\Theta \equiv \sum_{<a,b>} S_a S_b$ and the couplings
$K \equiv J/k_BT$. 
From Eq.~(\ref{expectval}) one gets
\begin{equation}
\langle M^2 \rangle_K = \frac{\sum_i{M_i^2 \exp(K \Theta_i)}}
{\sum_i{\exp(K \Theta_i)}}.
\label{momentofm2}
\end{equation}
Now, if one wants to calculate this same expectation value at some other 
temperature, one could store the histogram 
for $\Theta$ and then reweight it. 
But one can also write the desired expectation value 
\begin{equation}
\langle M^2 \rangle_{K^\prime} = \frac{\sum_i{M_i^2 \exp(K^\prime  \Theta_i)}}
{\sum_i{\exp(K^\prime \Theta_i)}}
= \frac{\sum_i{M_i^2 \exp(K \Theta_i) \exp((K^\prime - K) \Theta_i)}}
{\sum_i{\exp(K \Theta_i) \exp((K^\prime - K) \Theta_i)}},
\label{start}
\end{equation}
and expand the exponential containing $K^\prime - K$ in a Taylor 
series
\begin{equation}
\langle M^2 \rangle_{K^\prime} = 
\frac{\sum_i{M_i^2 \exp(K \Theta_i) 
\sum_{n=0}^\infty {(K^\prime - K)^n \Theta_i^n / n!}}}
{\sum_i{\exp(K \Theta_i) \sum_{n=0}^\infty {(K^\prime - K)^n \Theta_i^n / n!}}}.
\end{equation}
For finite systems the order of the sums can be safely interchanged, giving
\begin{equation}
\langle M^2 \rangle_{K^\prime} = 
\frac
{\sum_{n=0}^\infty{[(K - K^\prime)^n/n!]
\sum_i {M_i^2 \Theta_i^n \exp(K \Theta_i)}}} 
{\sum_{n=0}^\infty{[(K - K^\prime)^n/n!] 
\sum_i {\Theta_i^n \exp(K \Theta_i)}}}.
\end{equation}
One can now divide numerator and denominator by the partition
function $Z(K)$ and find that 
the terms at the end of both expressions are the expectation 
values (moments, in short), of $M^2 \Theta^n$ and $\Theta^n$, respectively,
taken at $K$.
Denoting $\delta K \equiv K - K^\prime$ one gets
\begin{equation}
\langle M^2 \rangle_{K^\prime} = \frac{\sum_{n=0}^{\infty} 
{(\delta K)^n \langle M^2 \Theta^n \rangle_K /n!}}
{\sum_{n=0}^\infty{(\delta K)^n \langle \Theta^n \rangle_K /n!}}.
\label{finish}
\end{equation}

Since the expansion of the exponential function is convergent everywhere,
there is no {\em a priori\/} limitation for this approach. 
But it is clear that one 
can never actually calculate all the required moments, and a truncation
needs to be applied. This immediately introduces very strong bounds
in the applicability 
of the method, since for large $\delta K$ the two sums in the 
previous expression require the inclusion of higher and higher moments
if reasonable results are going to be obtained.
A discussion about the number of moments needed to insure certain range
of convergence is given later on.

Once the series are truncated, one faces a more serious
difficulty: consider again the $M^2$ example,
and assume that a simulation has been conducted in a lattice with $N$ spins.
Since $M$ is an extensive quantity, 
$\langle M^2 \rangle$ is of order
$N^2$, and the range of $K$ where some degree of 
numerical convergence can be achieved is quite small. 
To find a way around this problem ---at least partially---, begin by working
with densities instead of with the original operators. 
Going back
to our example, 
define $m \equiv M/N$  and  $\theta \equiv \Theta/N$. 
The previous expression is rewritten as
\begin{equation}
\langle m^2 \rangle_{K^\prime} = \frac{\sum_{n=0}^{\infty} 
{(N \delta K)^n \langle m^2 \theta^n \rangle_K /n!}}
{\sum_{n=0}^\infty{(N \delta K)^n \langle \theta^n \rangle_K /n!}}
\end{equation}
Now it is clear that, since $\langle \theta \rangle$ is of order one,
one may expect numerical convergence of the {\em strongly truncated\/} series 
up to
\begin{equation}
\Delta K \approx \frac{1}{N} \quad \rightarrow \quad \Delta T \approx \frac{T^2}{N},
\label{rangenoshift}
\end{equation}
which is a 
very narrow range; here `strongly truncated' means a series with a few terms,
say, less than 20. This is actually a very conservative bound,
since the $n!$ in the denominators improve numerical convergence
(and guarantees it for the infinite series).

The convergence of the truncated series can be improved if one notices 
that the desired moments are taken from a sharply peaked distribution.
The combination of the fast increasing density of states $g(\Theta)$ 
and the fast decreasing Boltzmann weight implies the existence
of a narrow distribution $W_K(\theta)$ ---that is, the histogram in 
$\theta$---,
centered in some value $\theta_c$. Going back now to Eq.~(\ref{momentofm2}), 
one may see that 
the zero values of the operators appearing in the energy (and therefore in 
the energy density) can be easily shifted.
For instance, if one shifts the density $\theta$
by some value $\theta_R$, the expectation value for $m^2$
becomes
\begin{equation}
\langle m^2 \rangle_K = \frac{\sum_i{m_i^2 \exp(NK \theta_i)}}
{\sum_i{\exp(NK \theta_i)}}
= \frac{\sum_i{m_i^2 \exp(NK(\theta_i - \theta_R))}}
{\sum_i{\exp(NK(\theta_i - \theta_R))}},
\end{equation}
since the factors of $\exp(- N K \theta_R)$ cancel in 
the fraction. Now, if the 
value one chooses for $\theta_R$ is close to the center of the distribution, 
the
moments one gets for $\delta \theta \equiv \theta - \theta_R$ 
are going to be numerically small ---and
become smaller the higher the moment and the larger the lattice---, 
simply because of the narrowness
of the originating distribution.
Retracing the steps taken going from 
Eqs.~(\ref{start}) to~(\ref{finish}) one gets the final expression for 
$\langle m^2 \rangle_{K^\prime}$ as
\begin{equation}
\langle m^2 \rangle_{K^\prime} = \frac{\sum_{n=0}^{n^*} 
{(N \delta K)^n 
\langle m^2 (\delta \theta)^n \rangle_K /n!}}
{\sum_{n=0}^{n^*}{(N \delta K)^n \langle (\delta \theta)^n \rangle_K /n!}},
\label{finalexpansion}
\end{equation}
where both numerator and denominator series have been truncated to $n^*$ terms. 
Now the fast growth in the $N \delta K$ coefficients is partially balanced 
by a fast decrease in the values of $\langle (\delta \theta)^n \rangle_K$ 
and $\langle m^2 (\delta \theta)^n \rangle_K$, 
and in this way
the applicability of the method is extended to a much wider range in
couplings.
It is not too difficult to calculate the range of expected convergence
of the extrapolation. In fact, for temperatures away from the critical point
the width of $W_K(\theta)$ is order $1/\sqrt{N}$, from where one gets
a range of convergence for the extrapolation of 
\begin{equation}
\Delta K \approx \frac{1}{\sqrt{N}} \quad \rightarrow \quad \Delta T \approx \frac{T^2}{\sqrt{N}},
\label{rangenocritical}
\end{equation}
which comes from estimating $N \, \Delta K \, \Delta \theta \approx 1$.
Close to the critical point the width of $W_T(\theta)$
increases, and scales as 
$\Delta \theta \approx N^{\alpha/(2 \nu d)}/\sqrt{N}$.
From here one estimates then a range of convergence 
\begin{equation}
\Delta T \approx \frac{T^2}{\sqrt{N^{1 + \alpha/(\nu d)}}}.
\end{equation}
Here $\alpha$ and $\nu$ are the exponents for specific heath and correlation 
distance close to the critical point, and $d$ is the dimensionality. 
For a small $\alpha$ this is still a much wider range than the one 
gotten when no shifts are used. Notice that for the 2D Ising Model one replaces 
$N^{\alpha/(\nu d)}$ by $\ln N$ and gets an scaling
\begin{equation}
\Delta T \approx \frac{T^2}{\sqrt{N \ln N}}.
\label{rangecritical}
\end{equation}

Notice finally that the example given here, that extrapolates for
$m^2$, can be easily extended to any other lattice operator. 

\subsection{Example: the 2-D Ising Model: extrapolations at $h=0$}

A numerical test of the method was done for the two-dimensional
Ising Model in square lattices of sizes 16, 32 and 64, 
with the Hamiltonian given in Eq.~(\ref{ising2d}). 
Two Monte Carlo simulations were carried out using the Wolff algorithm~\cite{wolf1}, 
at a temperature of $k_B T/J  = 2.27$, close to the critical temperature for
the model. All moments of the form 
$\langle m^l (\delta \theta)^n \rangle$ and
$\langle (\delta |m|)^l (\delta \theta)^n \rangle$ for $0 \le l, n \le 16$ 
were recorded. First a short run with no shifts in either $\theta$ 
or $|m|$ was done, using 
$2 \times 10^6$, $3 \times 10^6$ and $4.4 \times 10^6$ Wolff 
iterations for $L=16$, $32$ and $64$, respectively, 
after discarding transients of $120$, $170$ and $270$ iterations.
The running time for this set of simulations was about $1,800$ seconds in 
a 2.4 GHz Pentium 4 processor.
The exact solutions for the energy and the specific heath were obtained
from the analytic free energy for finite lattices found by
Kaufman~\cite{kaufman1}.
Fig.~(1) shows the results of the 
extrapolation for the adimensional energy density $\langle \theta \rangle$ 
using either 14 or 15 moments~\cite{numberofmoments}, when no
shift in $\theta$ has been implemented, compared with the exact results. 
It is clear that the range of
applicability of the extrapolation is extremely narrow, 
as expected from Eq.~(\ref{rangenoshift}).
That estimate gives here $\Delta T_{32} = 0.0050$ and 
$\Delta T_{64} = 0.0013$. This estimate is quite conservative, and the
figure gives ranges of numerical convergence which are 2 to 3 times longer.

The second simulation was a larger run
where the results for $\langle \theta \rangle$ and
$\langle |m| \rangle$ from 
the first were used as reference values $\theta_R$ and $|m|_R$ for the shifts. 
The total numbers of Wolff iterations were 
$4 \times 10^7$, $6 \times 10^7$ and $8.8 \times 10^7$ 
for $L=16$, $32$ and $64$, respectively, after transients of 
$120$, $170$ and $250$ iterations were discarded. The total time for this set
of simulations was about $37,000$ seconds with the same processor.
Data were divided in 20 blocks in order to generate error estimates. 
Extrapolation for energy and specific heath were calculated.
The results
for the energies are given in Fig.~(2),
those for specific heath in Fig.~(3), and in both figures a
comparison with the exact results is given. It is
clear that the range of applicability of the extrapolation is
larger, and actually a bit larger than the estimation made in 
Eq.~(\ref{rangecritical}), which gives 
$\Delta T_{32} = 0.061$ and $\Delta T_{64} = 0.028$.
Two important points should be remarked: First, 
for $\langle \theta \rangle$ the direction in 
which the extrapolated curves deviate from the correct results, 
for low $T$, 
depends on the number of moments taken into account; 
they deviate upwards when one uses 
14 moments, and
downwards when using 15 moments. A similar behavior is found for $c$, 
except that now the two extrapolations deviate in different directions
{\em at both ends\/}
of the range of convergence. This gives a very simple way of 
bounding the
range of convergence of the algorithm. Second, the statistical errors in the 
extrapolations grow as one moves away from the simulated temperature, but
the effect of this growth is smaller than the effect of the change in 
the number of included moments. The behavior of both types of errors are 
given in Fig.~(4), which shows the difference between extrapolated and
exact energies for $L=32$. The behavior for errors is similar for the 
specific heath. Notice that the statistical errors in 
the extrapolated results actually become smaller for temperatures a 
bit below the point where the simulation was carried out; this curious 
phenomenon has been studied in full histogram 
extrapolations \cite{ferrenberg3}.

The moments of $|m|$ were used to generate an extrapolated estimation of
the susceptibility $\chi^\prime$, defined by 
$\chi^\prime = N K^\prime (\langle |m|^2 \rangle - \langle |m| \rangle^2) = 
N K^\prime (\langle m^2 \rangle - \langle |m| \rangle^2)$. 
This expression is used instead of the true susceptibility $\chi$
which in a numerical experiment does not give the expected peak close to the
critical temperature \cite{ferrenberg4}.
Fig.~(5) shows a comparison between the extrapolated $\chi^\prime$ and 
values obtained in other individual simulations. These were obtained at their
nominal temperatures using the Wolff algorithm.
It is remarkable
how the extrapolated results manage to reproduce reasonable well the 
peak in the susceptibility. Here one can also notice that the order of the
approximation again decides in which direction the extrapolated 
results deviate form the actual values, and so a simple comparison between 
the 15- and 16-moments expansions gives bounds for the region of
convergence of the method. 

\subsection{Example: the 2-D Ising Model with magnetic field}

As mentioned before, this algorithm becomes attractive especially in cases
where one has to deal with Hamiltonians composed of several 
operators, where the 
sizes of the histograms needed for reweighting may overflow the 
available memory. As a simple example consider
again the 2-dimensional nearest-neighbor
Ising Model, but now with a magnetic field. The Hamiltonian is 
\begin{equation}
{\cal H} =  -J \sum_{<a,b>} S_a S_b - H \mu \sum_a S_a, 
\label{ising2d-magnetic}
\end{equation}
giving a Boltzmann weight
\begin{equation}
\exp (-{\cal H}/k_B T)) = \exp(K \Theta + h M),
\end{equation}
where the additional definitions
$M \equiv \sum_a S_a$ and $h \equiv H \mu/k_BT$ have been introduced.
Consider now a simulation 
carried out at some temperature and magnetic field.
Assuming that one wants to extrapolate the 
expectation value of some operator $\Phi$,
one gets
\begin{equation}
\langle \Phi \rangle_{K^\prime, h^\prime} =
\frac
{\sum_i \Phi_i \exp(K \Theta_i + h M_i) 
\exp((K^\prime - K)\Theta_i + (h^\prime - h) M_i)}
{\sum_i \exp(K \Theta_i + h M_i) 
\exp((K^\prime - K)\Theta_i + (h^\prime - h) M_i)}.
\end{equation}
As before, it is better to change all operators into their densities,
defining $m \equiv M/N$ and $\phi \equiv \Phi/N$. 
Also, the density $\theta$ should be shifted so that its
distribution is centered close to zero (this is unnecessary for
$m$, since its distribution is symmetric around $m=0$). 
Performing these operations and the
Taylor expansions for the exponential containing $\delta K$ and $\delta h$,
one gets, repeating the same steps that gave 
Eq.~(\ref{finalexpansion}), the following expression
\begin{equation}
\langle \phi \rangle_{K^\prime h^\prime} = 
\frac
{ \sum_{l=0}^{\infty} \sum_{n=0}^{\infty} 
(N \delta K)^l (N \delta h)^n 
\langle \phi (\delta \theta)^l m^n \rangle_{K,h} /(l! n!)}
{ \sum_{l=0}^{\infty} \sum_{n=0}^{\infty} 
(N \delta K)^l (N \delta h)^n 
\langle (\delta \theta)^l m^n \rangle_{K,h} /(l! n!)}.
\end{equation}
One should pay attention here to the fact that for low temperatures 
the histogram $W_{K,h}(\theta, m)$ becomes bimodal in $m$, and the assumption
of a distribution with a single narrow peak is no longer valid. However, 
close to the critical point the width in $m$ of such histogram
remains small, and one can still get by with the first few moments.  
Occasionally, it may be convenient to work in terms of $|m|$, whose 
distribution remains unimodal.

The data obtained from the previous simulations were now used to generate the
behavior of the magnetization and the susceptibility 
at non-zero values of $H$. 
For the magnetization one gets, after truncation
\begin{equation}
\langle m \rangle_{K^\prime h^\prime} = 
\frac
{ \sum_{l=0}^{l^*} \sum_{n=0}^{n^*} 
(N \delta K)^l (N h)^n 
\langle (\delta \theta)^l m^{n+1} \rangle_{K,h=0} /(l! n!)}
{ \sum_{l=0}^{l^*} \sum_{n=0}^{n^*} 
(N \delta K)^l (N h)^n 
\langle (\delta \theta)^l m^n \rangle_{K,h=0} /(l! n!)};
\label{two_pars}
\end{equation}
an analogous expression is obtained for 
$\langle m^2 \rangle_{K^\prime, h^\prime}$, and from here the true
susceptibility can be computed.
The results are shown in Fig.~(6), which shows: (a) magnetization 
vs.\ $H$ for $L=16$, $32$ and $64$ at the critical temperature $T_c = 2/\ln (1 + \sqrt{2}) = 
2.269185...$, and (b) magnetization vs.\ $H$ for 
$L= 32$ at $T=2.20$, $2.27$ and
$2.34$. In all cases the magnetization has been extrapolated from the
simulations that were carried out at $H=0$ and $T=2.27$.
In a slight departure from what was done before, here the 
denominator in Eq.~(\ref{two_pars}) was calculated using 16 moments while 
the sums in the numerator were truncated to $15$ or $14$ moments.
The individual points have been calculated using a 
modified version of the Wolff algorithm, were the acceptance ratio depends on
the change of energy due to a cluster flipping in the presence a 
magnetic field $H$. It is clear from the figure
that the expansion in moments
reproduces quite well the behavior of the magnetization for each
temperature and lattice size,
and in particular it manages to show the large growth of $m$ with $H$
as $T$ is reduced. The splits in the extrapolation curves correspond
to the separation of the $15$- and $14$-moment extrapolations, and mark
the end of the ranges of convergence. For small lattices these
splits do not appear in the $H$ range tested here.

Finally, Figs.~(7) and~(8) show the true susceptibility $\chi$ as a 
function of $H$,
for $T= T_c$ and $L=16$, $32$ and $64$, and for $L=32$ and $T=2.20$,
$2.27$ and $2.34$.
It should be noticed that the extrapolation manages to cover quite well
the whole peak in $\chi$. The results for $L=32$ and different temperatures
also show an excellent agreement between extrapolations and individual
simulations, and show how the method can really
extrapolate in more than one parameter. Notice that, as expected,
$\chi$ grows as $T$ is reduced, even below $T_c$; for
$H=0$ the true susceptibility is just proportional to $\langle m^2 \rangle$.
Otherwise, the behavior
the extrapolated quantities {\em vis a vis\/} the individual simulations 
at $H \ne 0$ is in all respects analogous the results found before:
a very good correspondence for small $H$ 
---that is, close to where the simulation
was carried out---, followed by a large deviation, 
which depends on the number of moments included in the extrapolation.

\section{Conclusions}

This paper shows how to implement the histogram method for extrapolation of 
results of a Monte Carlo simulation using the moments of the histogram. 
This approach has several advantages over the direct method ---histogram
construction and posterior reweighting---, and over the method of 
storing configurations for their reweighting (named ``histogram on the fly'' 
in Ref.~\cite{ferrenberg3}). To start with, the resulting expressions for the 
extrapolated quantities are given by very simple and conceptually
appealing formulas. 
Second, the ranges of applicability of the method become evident simply 
by changing the number of 
moments included in the extrapolation. Third, the amount of 
computer memory and physical storage needed are so small that one may 
without any problem generate several repetitions of the simulations so 
as to generate in a simple way the error estimates for the extrapolated
quantities. And finally, this approach eliminates completely the need to 
choose a binning size in cases of continuous variables.

One needs however to balance these benefits against the cost of the 
extra approximation involved in the method. After all, replacing a
full histogram for its first few moments necessarily reduces precision. 
How many moments one really needs to keep in any given simulation 
so that not too much information is lost is an issue that has to be considered
with some care.
On the one hand, it is clear that increasing too much the number of 
stored moments not only defeats one of the motivations for this approach,
which is to work with a limited memory, but also necessarily runs into 
the limits of reliability imposed by the statistical errors of the simulation. 
On the other, not keeping enough moments implies a waste of simulation time.
As a first approach to the answer to this question one can consider
the following estimation, done here for a one-coupling Hamiltonian: consider the 
preliminary run needed for the estimation of $\theta_c$. This preliminary run 
can also be used to obtain a rough estimate of the width $\sigma_\theta$ 
of the distribution. Now, the order of magnitude of the moments 
$\langle (\delta \theta)^n \rangle$ will be around 
$\sigma_\theta^n$, and so the 
terms needed in an estimation with a shifted coupling look like 
(see Eq.~(\ref{finalexpansion})) 
\begin{equation}
\frac{(\Delta K)^n \langle (\delta \theta)^n \rangle}
{n!}
\approx
\left( \frac{e \Delta K \sigma_\theta}{n} \right)^n.
\end{equation}
It is not difficult to show, using a steepest descent approximation,
that as a function of $n$ this expression 
behaves as a Gaussian centered in $n_{\max} = \Delta K \, \sigma_\theta$, with a
width given by $\sqrt{n_{\max}}$ and height $\exp(n_{\max})$. Therefore one gets
\begin{equation}
\frac{(\Delta K)^n \langle (\delta \theta)^n \rangle}
{n!}
\approx
\exp(\Delta K \sigma_\theta) \, \exp \left( \frac
{(n - \Delta K \sigma_\theta)^2}{2 \Delta K \sigma_\theta} \right).
\end{equation}
The conclusion is then the following: given an initial estimation of the width 
$\sigma_\theta$ of the distribution, and assuming that a certain maximum 
extrapolation
range $\Delta K$ is going to be used, the main contribution to 
the extrapolation comes
from the moments with $n$ around $n_{\max} = \Delta K \sigma_\theta$. Besides,
one finds that the moments 
with $n$ such that $n - n_{\max} \gg \sqrt{n_{\max}}$ are basically irrelevant.
The numerical results shown here display a much larger dependence on the number of 
moments used than on the statistical spread of the data, suggesting 
that several more moments may have been used in the extrapolation 
before reaching the limits given by statistical spread.

\acknowledgments{I want to thank F.\ Sastre for a careful reading of the manuscript 
and for many helpful comments and suggestions.
This work has been supported by CONACyT through grant No. 40726-F.}

\begin{figure}
\includegraphics[width=17.0cm]{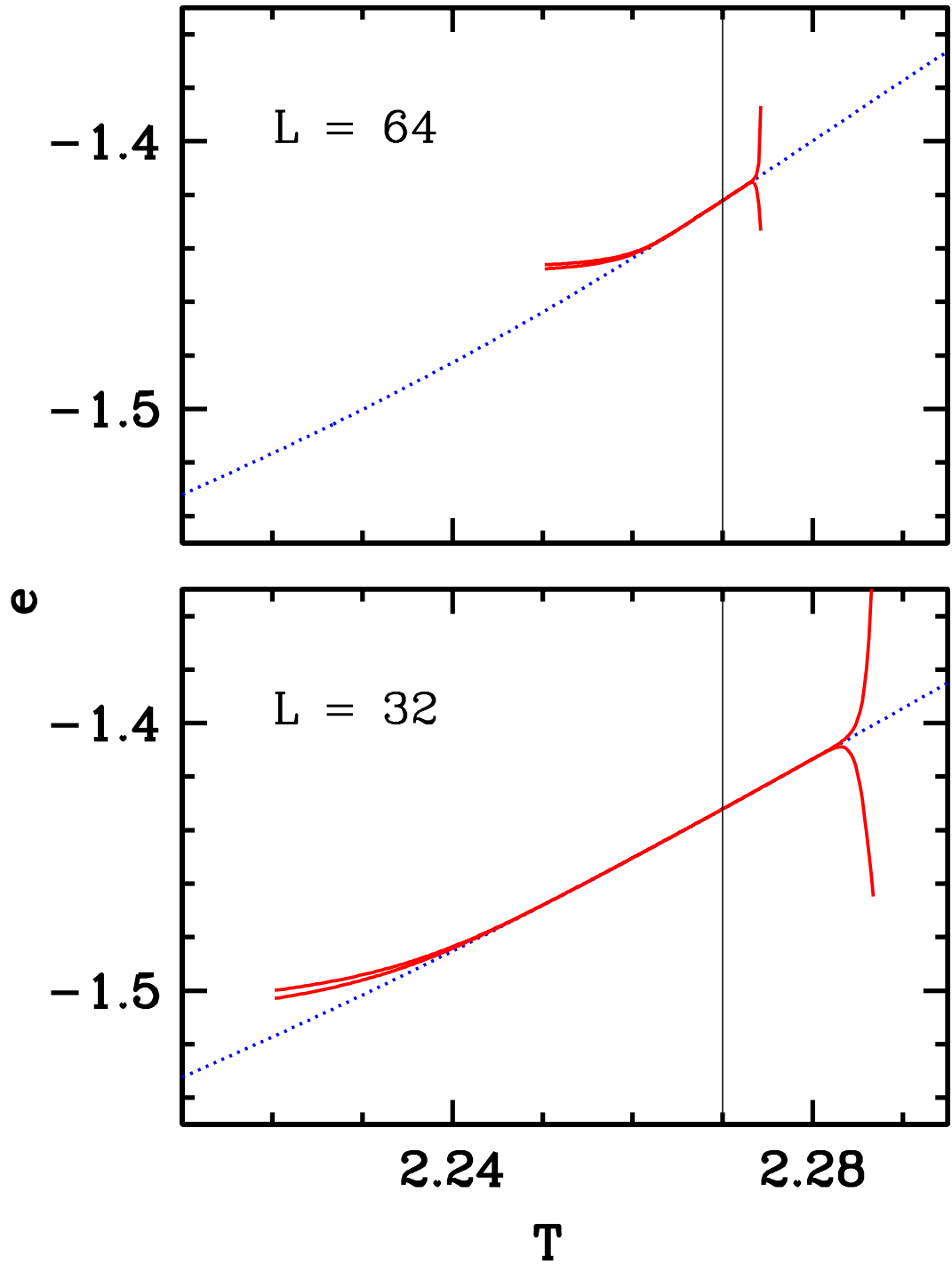}
\caption{
Energy per site vs.\ Temperature for the 2-D Ising 
Model, for $L=32$ and $L=64$. The dotted line gives the exact results 
obtained from the Kauffman solution. 
The solid lines give 14- and 15-moment extrapolations using values from 
a simulation at $T=2.27$ (indicated by the vertical line). 
Here the moments are taken without using any shifts ($\theta_R=0$).}
\label{f1}
\end{figure}

\begin{figure}
\includegraphics[width=17.0cm]{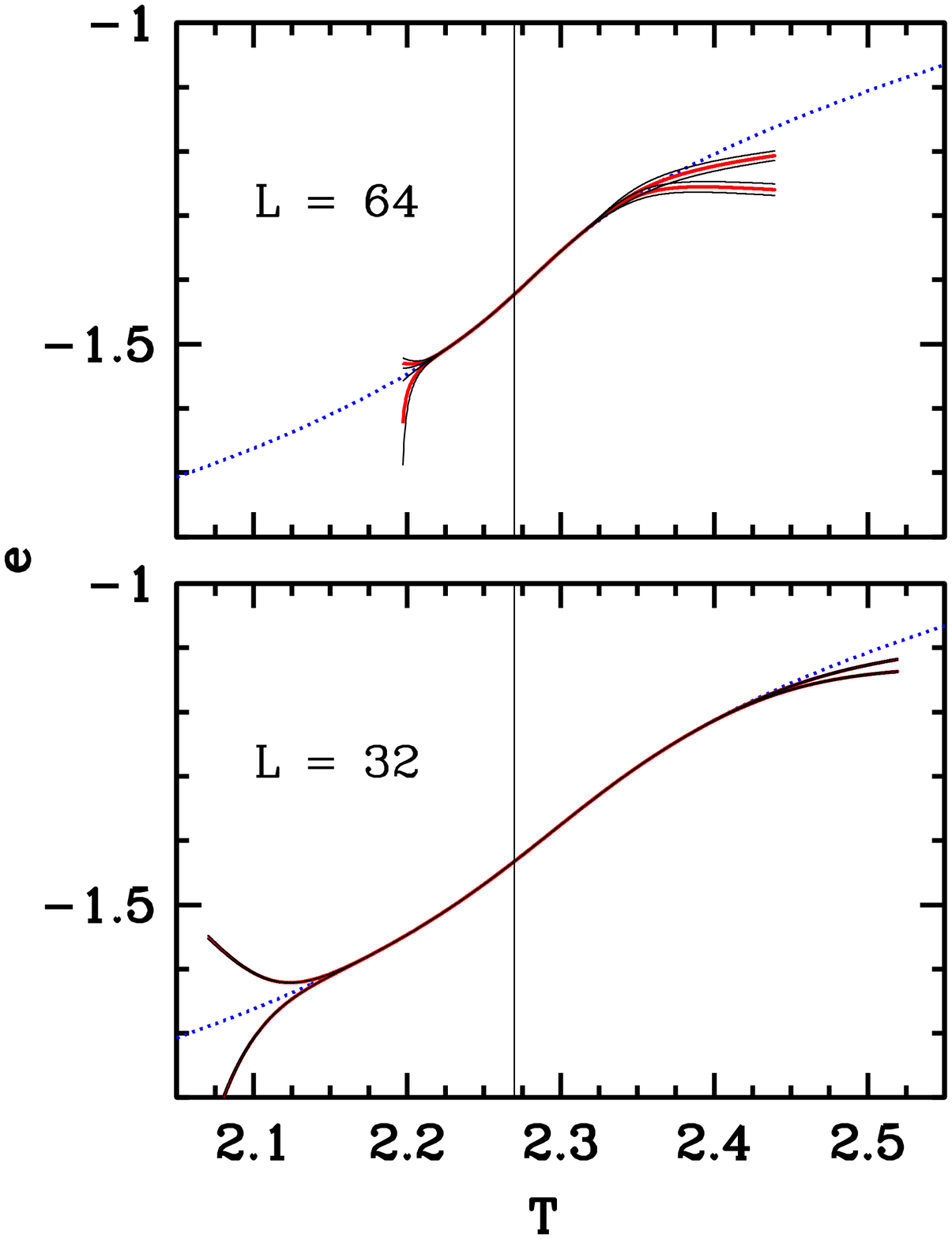}
\caption{
Energy per site vs.\ Temperature for the 2-D Ising 
Model, for $L=32$ and $L=32$. The dotted line gives the exact results
obtained from the Kaufman solution.
The thick solid lines give 14- and 15-moment extrapolations using values 
from a simulation at $T=2.27$ (indicated by the vertical line). The
thin solid lines give the error bars for the extrapolations. 
Here the moments are taken using values of $\theta_R$ 
of $-1.432094$ for $L=32$ and 
$-1.422108$ for $L=64$.}
\label{f2}
\end{figure}

\begin{figure}
\includegraphics[width=13.0cm,angle=270]{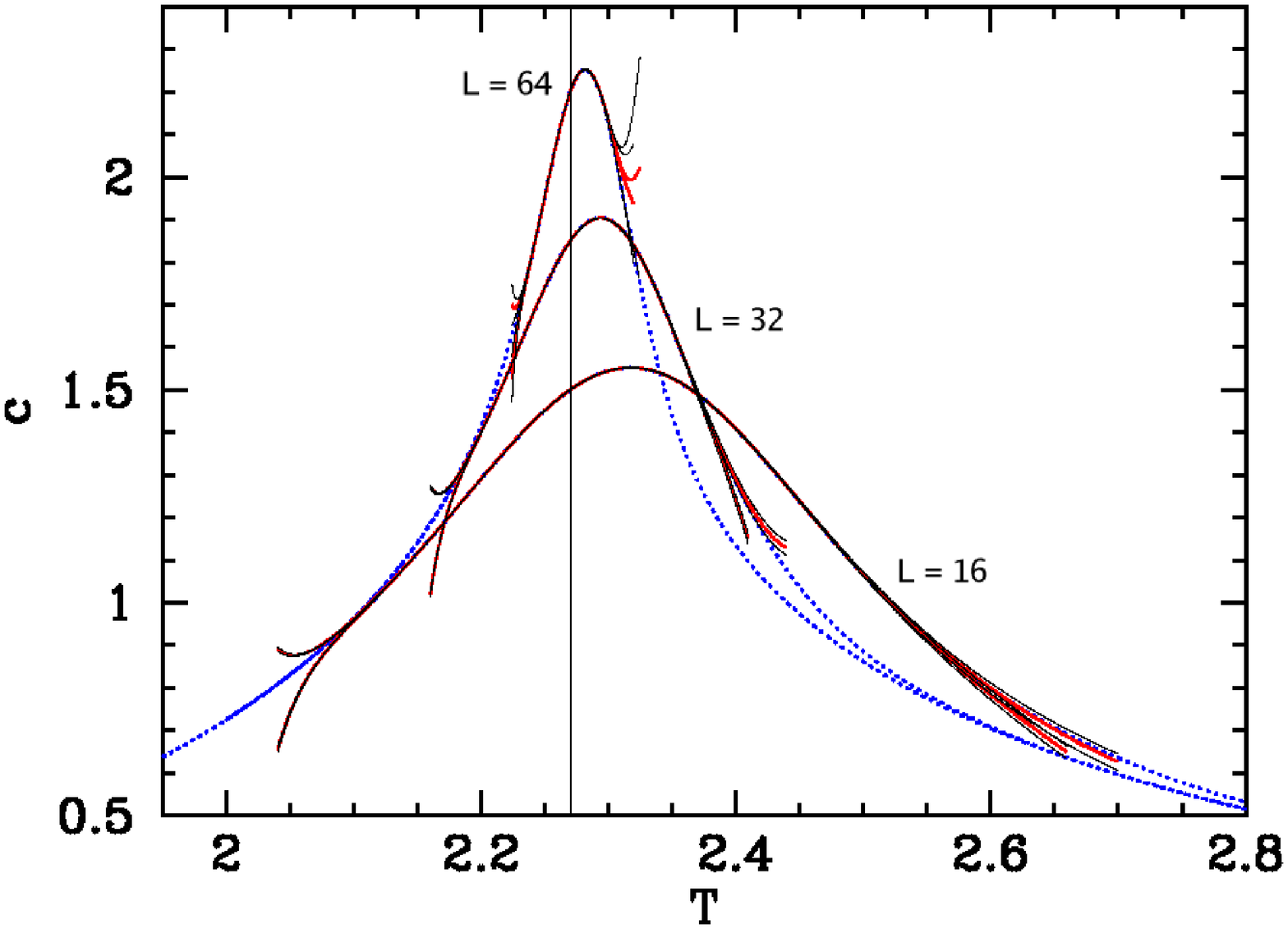}
\caption{
Specific heath per site vs.\ Temperature for the 
2-D Ising Model, for $L=16$, $32$ and $64$. The dotted lines give the 
exact results obtained from the Kaufman solution.
The thick solid lines give 13- and 14-moment extrapolations using values 
from a simulation at $T=2.27$ (indicated by the vertical line). The
thin solid lines give the error bars for the extrapolations. 
Here the moments are taken using values of $\theta_R$ 
of $-1.451724$ for $L = 16$,
$-1.432094$ for $L=32$, and $-1.422108$ for $L=64$.}
\label{f3}
\end{figure}

\begin{figure}
\includegraphics[width=13.0cm,angle=270]{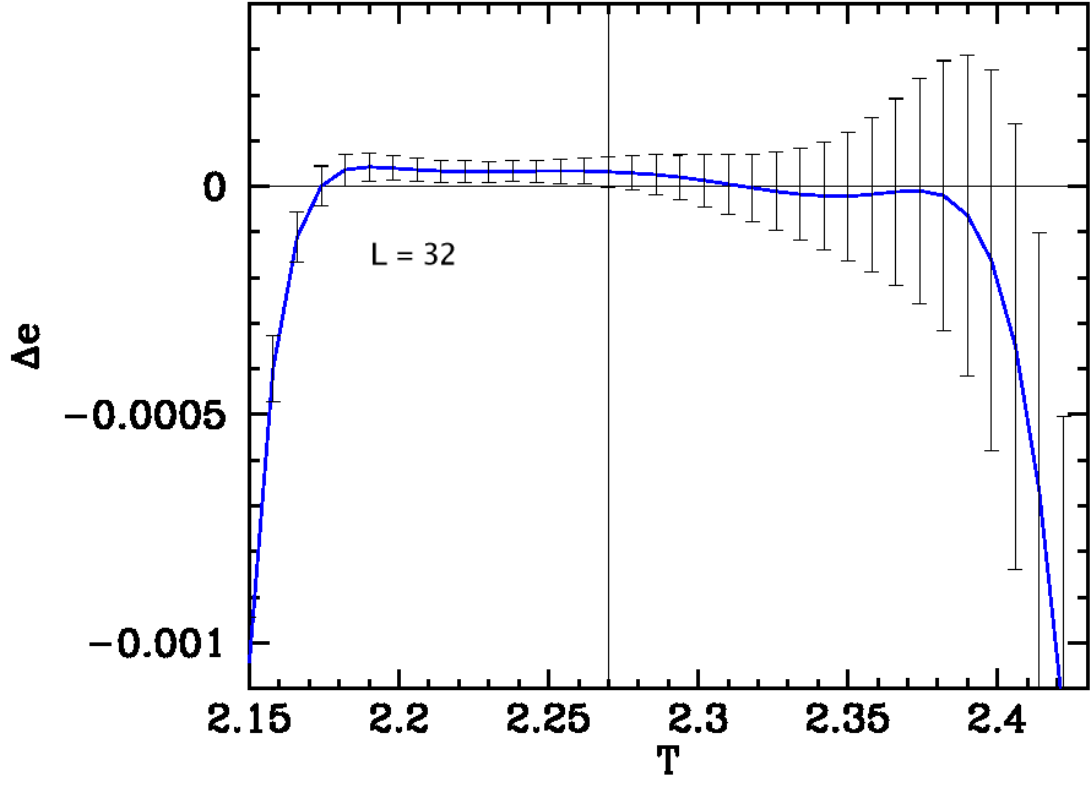}
\caption{
Error in the extrapolation of the energy density
for $L=32$. The solid line gives the difference between the extrapolated
energy and the exact Kaufman solution. The error bars correspond
to the extrapolation. Here 15 moments were used. Here $\theta_R = -1.432094$.}
\label{f4}
\end{figure}

\begin{figure}
\includegraphics[width=13.0cm,angle=270]{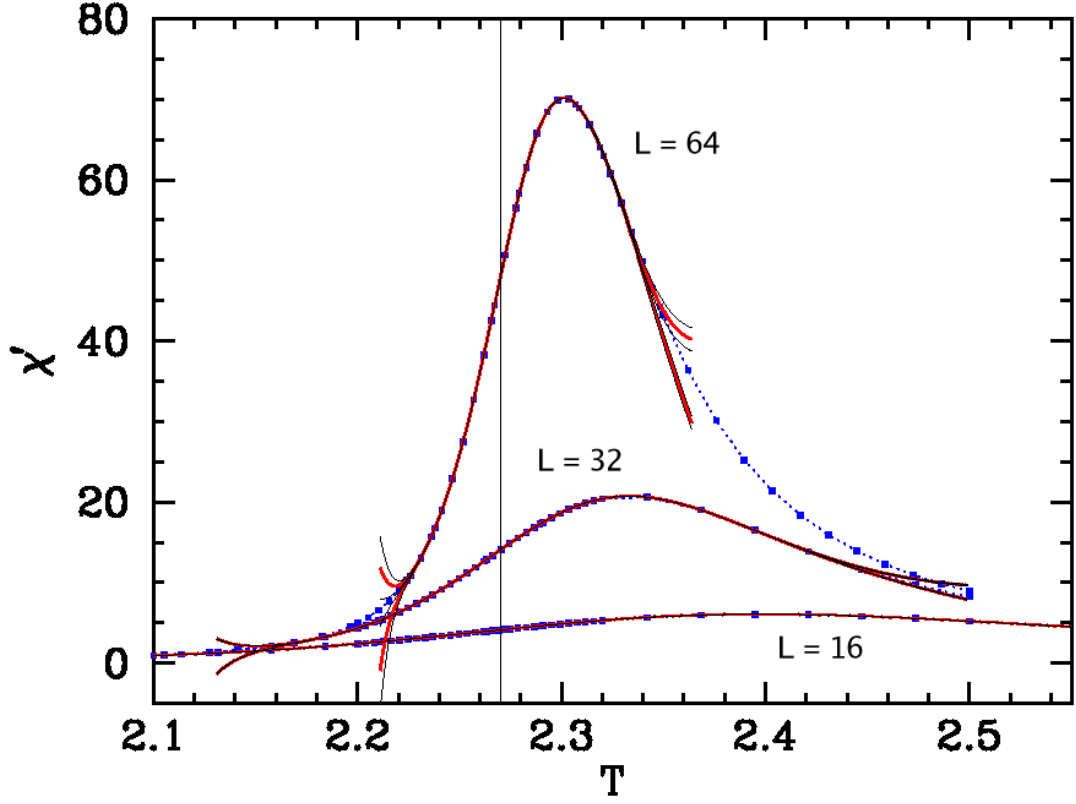}
\caption{
Susceptibility $\chi^\prime$, calculated using 
$|m|$, for the 2-D Ising model with $L=16$, $32$ and $64$. The points are
from independent simulations, carried out at their nominal temperatures, 
an the dotted line is given only as a guide to the eye. The solid 
lines correspond to extrapolations from a simulation at $T=2.27$, using
15 and 16 moments.}
\label{f5}
\end{figure}

\begin{figure}
\includegraphics[width=13.0cm,angle=270]{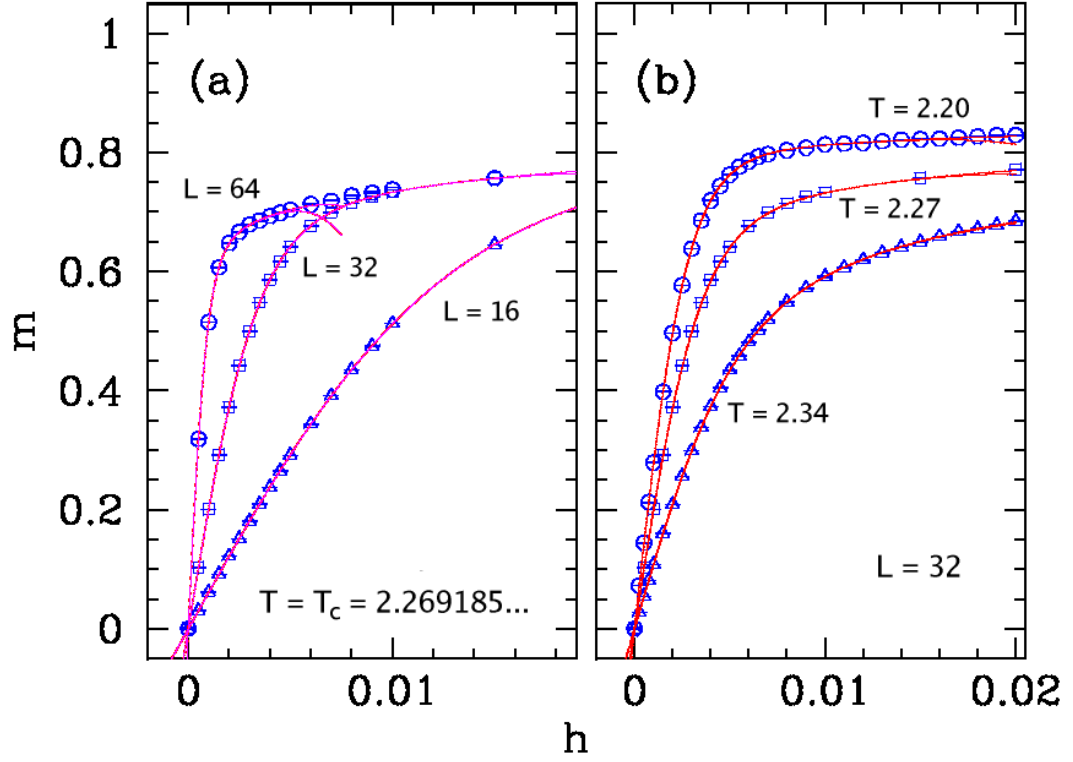}
\caption{
Magnetization vs.\ $H$ for the 2-D
Ising model. The lines give the values extrapolated from a 
simulation at $T=2.27$, 
$H=0$, and the points, with their error bars,
correspond to individual simulations carried out
at their nominal $T$ and $H$ values.
(a): Results for $T=T_c = 2.269185\ldots$, with sizes $L=16$ (triangles), 
$32$ (squares) and $64$ (round marks). The split visible for 
the $L=64$ line correspond to the separation between the 14- and the 
15-moments extrapolations. 
(b) Results for $L=32$, with temperatures 
$T=2.34$ (triangles), $T=2.27$ (squares)
and $T=2.20$ (round marks). 
In both cases the extrapolated statistical errors come 
out smaller than the thickness of the lines.}
\label{f6}
\end{figure}

\begin{figure}
\includegraphics[width=13.0cm,angle=270]{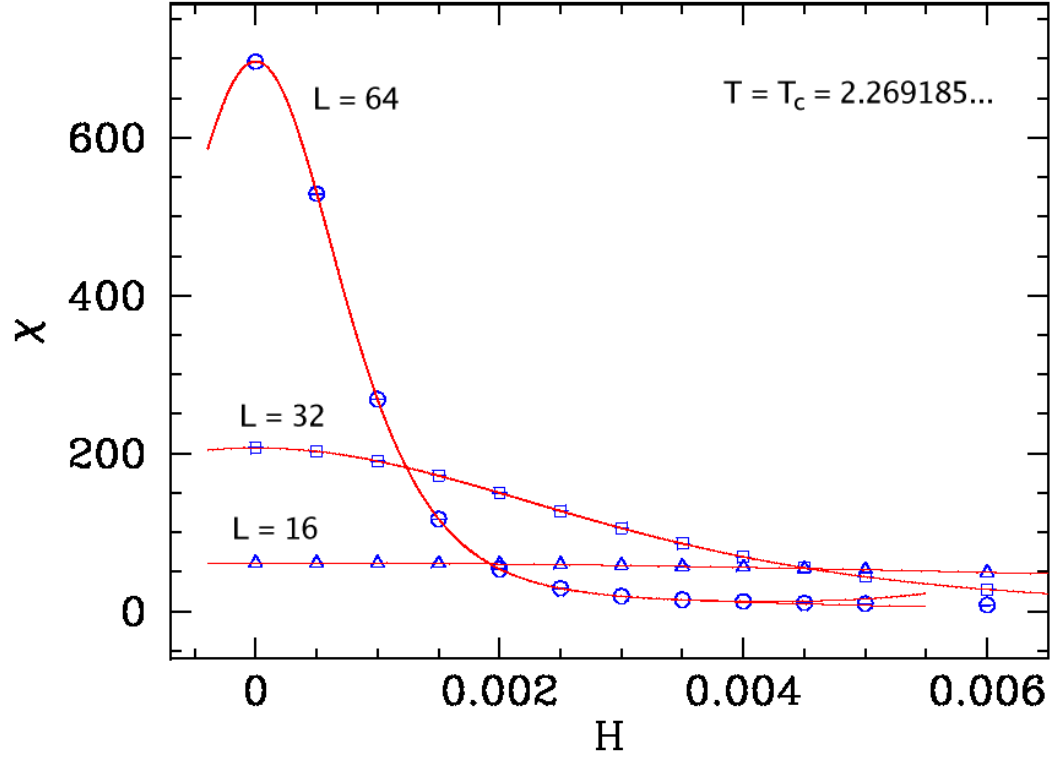}
\caption{
True susceptibility $\chi$ vs.\ $H$ for the 2-D
Ising model. The lines give the values extrapolated from a 
simulation at $T=2.27$, 
$H=0$, and the points, with their error bars,
correspond to individual simulations carried out
at their nominal $T$ and $H$ values.
Results shown here are for $T=T_c = 2.269185\ldots$, with sizes 
$L=16$ (triangles), $32$ (squares) and $64$ (round marks). 
The split visible for the $L=64$ line correspond to the separation 
between the 14- and the 15-moments extrapolations. 
The extrapolated statistical errors come 
out smaller than the thickness of the lines.}
\label{f7}
\end{figure}

\begin{figure}
\includegraphics[width=13.0cm,angle=270]{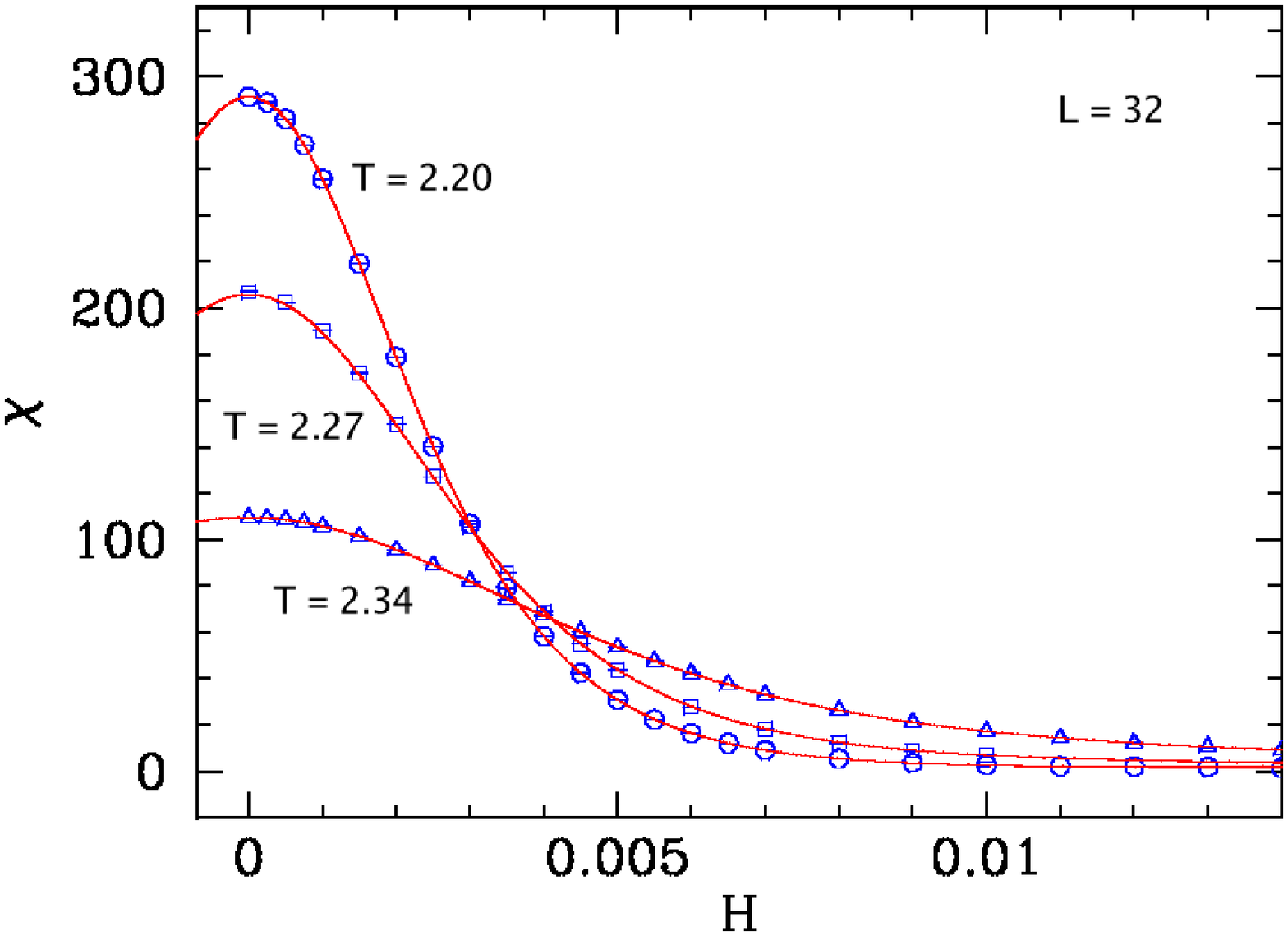}
\caption{
True susceptibility $\chi$ vs.\ $H$ for the 2-D
Ising model. The lines give the values extrapolated from a 
simulation at $T=2.27$, 
$H=0$, and the points, with their error bars,
correspond to individual simulations carried out
at their nominal $T$ and $H$ values.
Results shown here are for $L=32$, with temperatures 
$T=2.34$ (triangles), $T=2.27$ (squares)
and $T=2.20$ (round marks). 
The extrapolated statistical errors come 
out smaller than the thickness of the lines.}
\label{f8}
\end{figure}


\begin{thebibliography}{99}

\bibitem[*]{gp} Electronic address: gperez@mda.cinvestav.mx

\bibitem{metropolis1} N.\ Metropolis, A.\ W.\ Rosenbluth, N.\ M.\ Rosenbluth,
A.\ M.\ Teller and E.\ Teller, J.\ Chem.\ Phys. {\bf 21}, 1087 (1953).

\bibitem{newman-barkema} See, e.g., M.\ E.\ J.\ Newman and G.\ T.\ Barkema,
{\em Monte Carlo Methods in Statistical Physics\/}, Oxford U.\ P.\
(Oxford, 1999).

\bibitem{landau-binder} See, e.g., D.\ P.\ Landau and K.\ Binder, 
{\em A Guide to Monte Carlo Simulations in Statistical Physics\/}, 
Cambridge U.\ P. (Cambridge, 2000)

\bibitem{ferrenberg1} A.\ M.\ Ferrenberg and R.\ H.\ Swendsen, 
Phys.\ Rev.\ Lett. {\bf 61}, 2635 (1988); 
Phys.\ Rev.\ Lett. {\bf 63}, E1658 (1989).

\bibitem{ferrenberg2} A.\ M.\ Ferrenberg and R.\ H.\ Swendsen, 
Phys.\ Rev.\ Lett. {\bf 63}, 1195 (1989).

\bibitem{valleau1} A large list of works in the 
Histogram Method that predate the papers of Ferrenberg and Swendsen 
is given in Ref.~[18] of J.\ Lee and J.\ M.\ Kosterlitz, Phys.\ Rev.\ B
{\bf 43}, 3265 (1991).

\bibitem{not-dense} It may be argued that there is no need for such a 
large array to be stored, since in practice many bins will never get
hit because the corresponding combination of values for the three operators
(nearest-neighbor, next-nearest-neighbor and magnetic) is not allowed.
This observation
is correct, but in general it may be quite difficult to know
beforehand which combinations of lattice operators 
are allowed and which are not, or even how many 
are allowed. It is then not possible to request the correct amount of
computer memory at the start of a simulation.
A possible solution is to request new memory locations only when needed,
but then one is never sure about the maximum lattice size
that can be safely explored.

\bibitem{ferrenberg3} A.\ M.\ Ferrenberg, D.\ P.\ Landau and R.\ H.\ Swendsen, 
Phys.\ Rev.\ E {\bf 51}, 5092 (1995).

\bibitem{rickman1} J.\ M.\ Rickman and S.\ R.\ Phillpot, 
Phys.\ Rev.\ Lett. {\bf 66}, 349 (1991).

\bibitem{wolf1} U.\ Wolff, Phys.\ Rev.\ Lett. {\bf 62}, 361 (1989).

\bibitem{kaufman1} B.\ Kaufman, Phys.\ Rev. {\bf 76}, 1232 (1949).

\bibitem{numberofmoments} Since the numerator in the extrapolation of 
$\langle \delta \theta \rangle$ contains moments of the form 
$\langle (\delta \theta)^{n+1} \rangle$, one can run $n$ up to $15$ when $16$ moments 
are kept. For simplicity, the same bound $n^*$ is used for the
denominator. Similarly, for the extrapolation of $\langle (\delta \theta)^2 \rangle$ 
one can run $n$ up to $14$ when $16$ moments 
are kept.

\bibitem{ferrenberg4} A.\ M.\ Ferrenberg and D.\ P.\ Landau, 
Phys.\ Rev.\ B {\bf 44}, 5081 (1991).

\end{thebibliography}
\end{document}